\documentstyle[11pt,newpasp,twoside, epsf]{article}
\def\edcomment#1{\iffalse\marginpar{\raggedright\sl#1\/}\else\relax\fi}
\reversemarginpar

\begin{document}
\title{SACY - Present Status \altaffilmark {1} }
\altaffiltext {1}{Based on observations made under the
Observat\'{o}rio Nacional-ESO agreement for the joint operation of
the 1.52\,m ESO telescope and at the  Observat\'{o}rio do Pico dos Dias,
operated by MCT/Laborat\'{o}rio Nacional de Astrof\'{\i}sica,  Brazil}
\author{ C. A. O. Torres, G.R. Quast}
\affil{Laborat\'{o}rio Nacional de Astrof\'{\i}sica/MCT, 37504-364
Itajub\'{a}, Brazil}
\author{R. de la Reza,  L. da Silva}
\affil{Observat\'{o}rio Nacional/MCT, 20921-030 Rio de Janeiro, Brazil}
\author{C. H. F. Melo}
\affil{ESO, Casilla 19001 Santiago 19, Chile}

\begin{abstract}

The scientific goal of the SACY (Search for Associations
Containing Young-stars) project is to identify eventual associations
of stars younger than the Local Association,
spread among the optical counterparts of the
ROSAT X-ray bright sources.
High-resolution spectra for possible optical counterpart
later than G0 belonging to HIPPARCOS and/or TYCHO-2 catalogues
were obtained in order to assess both the
youth and the spatial motion of each target.
The newly identified young stars present a patchy distribution
in UVW space and in the sky as well
revealing the existence of huge nearby young associations.
Here we present the associations identified in the present sample.

\end{abstract}

\section{Introduction}
The detection of X-ray sources  by the ROSAT All-Sky Survey (RASS)
associated with TTS outside star formation
regions (Neuh\"{a}user 1997) gave a tool to find new young associations.
In fact, Torres et al. (2000), and Zuckerman \& Webb (2000),
using these sources, found evidences for two young associations
near the South Celestial Pole, the Horologium (HorA) and the Tucana (TucA) Associations.
To examine the possibility that these associations are the same and
to search for other ones we undertook a Search for Associations Containing
Young-stars (SACY) (de la Reza et al. 2001; Torres et al. 2001;
Quast et al. 2001).
In the SACY we selected and observed all bright RASS sources that
could be associated with
TYCHO-2 or HIPPARCOS stars with (B-V) $>$ 0.6, excluding very well known
RS CVn, W UMa, giants, etc in  SIMBAD.

We obtained high resolution spectra for the selected candidates with
FEROS echelle spectrograph (Kaufer et al. 1999)
(resolution of 50000; spectral coverage of 5000\,\AA)
of the 1.52 m ESO telescope at La Silla
or  with the coud\'{e} spectrograph
(resolution of 9000; spectral coverage of 450\,\AA, centered at 6500\,\AA)
of  the 1.60\,m telescope of the Observat\'{o}rio do Pico dos Dias.
For some stars we obtained
radial velocities with  CORALIE at the Swiss Euler Telescope at ESO
(Queloz et al. 2000).

>From  the collected spectra we have obtained spectral classifications,
radial velocities and equivalent widths of Li\,I lines.
In particular, the Li\,I line is important since it can provide
a first age estimate (Jeffries 1995) for late type stars
allowing us to select possible Post-T Tauri stars.
For a given star, if its Li\,I resonance line equivalent width is located
near or above the Li\,I line delimited by the members of the Local
Associations clusters (Neuh\"{a}user 1997), it is  flagged as
young star.

\section{Results}

There are 9574 ROSAT bright sources in the Southern Hemisphere,
2071 of them with \bv $>$ 0.6 in TYCHO-2.
Until the La Serena meeting (2002, March)
we observed 687 stars, and for 170 more stars we got
relevant data from the literature.
>From all these stars, 283 fulfill our
definition for a young star (239 from our spectroscopic observations).

As most of these young stars have no HIPPARCOS parallaxes, we
applied the following kinematical analysis to find possible associations:
Each point in UVW velocity space  is taken as a convergence point and
we calculated for it the parallaxes of all stars such as to minimize the
moduli of the space velocity vectors relative to this point
(but, of course, preserving the parallaxes of HIPPARCOS stars).
Then we calculated the density of stars in the velocity space around
each point of a grid in UVW.
Around some points there are density concentrations much larger than
the background fluctuation, revealing possible associations. 
All the main concentrations are also constricted in space, but
some of them cover large areas in the sky.

\section{The Young Associations}

In Tables 1  \& 2 we present the properties of the eight young
associations we found until now.
For those that engulf previously known ones, we use
bonafide members (not observed in the SACY) to help in their definition. 
The quantity of bonafide members used are
indicated as  the last number in Table 1. 
In Table 2 we indicate the distance ($\rho_{max}$)
of the farthest member to the calculated center of
the association, giving an idea of its size.
It should be noted that the method gives no unique solution for
some stars, but in almost all cases there is an obvious better
membership.
As there are important areas not covered until now by the SACY,
and specially the Sco-Cen Association region, these identifications
are preliminary.

\begin{figure}

\
\begin{center}
{\leavevmode
\epsfxsize=.85\textwidth \epsfbox{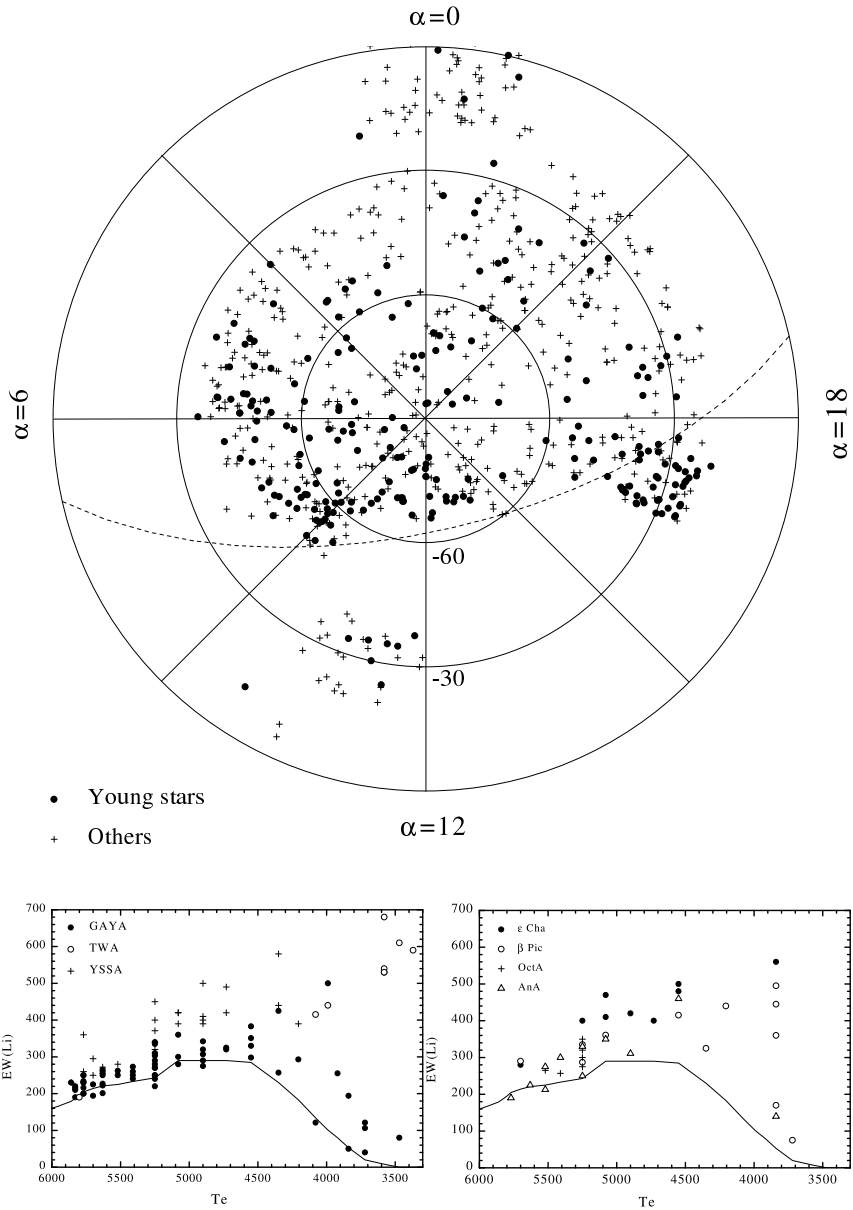}}
\label{fig:figura}
\end{center}

\caption{{\bf Top}. Celestial polar projection of the stars observed in SACY.
Stars classified as young according to their Li\,I equivalent width are plotted as
filled circles. More evolved stars are indicated as crosses.
{\bf Bottom left}. Li\,I equivalent width as a function of the effective temperature
for the members of GAYA
(filled circles), TWA (open circles) and YSSA (crosses).
The continuous line indicates the upper limit of the
Li\,I equivalent width observed in the members of the Local Association clusters.
{\bf Bottom right}.
The same as bottom
left for $\epsilon$ Cha (filled circles), $\beta$ Pic (open circles), OctA (crosses) and
AnA (open triangles).}
\end{figure}

\begin{table}
\caption{Space motions and parallaxes of SACY's Young Associations}
\begin{tabular}{lrrrrrrrrr}
\tableline

Name&U&$\sigma$&V&$\sigma$&W&$\sigma$&$\pi$&$\sigma$&N$_*$\\
&km/s&km/s&km/s&km/s&km/s&km/s&mas&mas\\
\tableline
GAYA1&-9.8&0.9&-21.7&1.0&-0.9&0.9&16.3&2.2&31\\
GAYA2&-10.9&0.9&-22.4&0.8&-4.7&1.1&11.8&4.4&36\\
TWA&-10.8&1.2&-17.7&1.1&-5.6&1.2&21.2&8.3&5+5\\
YSSA&-3.7&1.4&-13.5&1.1&-8.2&1.3&8.5&1.9&20+8\\
$\epsilon$ ChaA&-7.9&0.5&-18.9&0.6&-7.7&1.0&11.4&0.6&8+2\\
$\beta$ PicA&-9.3&1.3&-16.1&0.8&-8.8&0.9&29.9&24.5&11+2\\
OctA&-9.9&0.5&-1.6&0.6&-7.3&0.4&9.7&1.3&6\\
AnA&-8.5&0.7&-28.7&0.6&-10.1&1.1&11.6&7.9&9\\
\tableline
\tableline
\end{tabular}

\end{table}
\begin{table}
\caption{Positions relative to the Sun}
\begin{tabular}{lrrrrrrrr}
\tableline

Name&X&$\sigma$&Y&$\sigma$&Z&$\sigma$&$\rho_{max}$&N$_*$\\
&pc&pc&pc&pc&pc&pc&pc&\\
\tableline
GAYA1&18&22&-57&41&-30&10&88&31\\
GAYA2&8&25&-77&34&-38&24&88&36\\
TWA&14&13&-49&24&21&7&50&10\\
YSSA&118&22&-8&9&-10&27&63&28\\
$\epsilon$ ChaA&44&4&-74&6&-16&6&14&10\\
$\beta$ PicA&37&32&-9&16&-15&6&52&13\\
OctA&59&25&-67&7&-50&4&42&6\\
AnA&74&68&-46&46&-32&36&130&9\\
\tableline
\tableline
\end{tabular}
\end{table}

The young association found up to now are the following:

{\it a) GAYA} - The Great Austral Young Association (Torres et al. 2001) seems to
split in two parts, although it is not clear if both are actually distinct
associations or if this split was created by some bias or by our method.
Their separation is mainly in W velocity and in distance.
The previous HorA and TucA are within GAYA.
Some of the proposed members  of TucA are outside of
the velocities definition (mainly the eastern ones).
Until now both GAYAs seem deficient in binaries.
There is a proposed SB2  member for GAYA2 but a radial velocity
search on 28 members (more than a third of the members) has give no
other SBs.
The western proposed members are near the present survey limit.

{\it b) TWA} - Until now we have found no new members.
Of the known ones only five fulfill the SACY criteria.
Our kinematical solutions put some stars farther away.
TWA12 and TWA19 would be at about 100pc.

{\it c) YSSA} - There is a group of young stars, spread from $\rho$ Oph to R Cra,
with very similar properties, that we are now calling the
Young Sco-Sgr Association.
As we have no data yet for the Upper Sco subgroup we can not be sure if
YSSA belongs or not to it.
The eastern border of YSSA is at one of the present limits of SACY.
The western border engulfs the stars mentioned in  Quast et al. (2001) and   Neuh\"{a}user  et al. (2000).
Note the very well defined distance, near the assumed for the R CrA clouds.

{\it d) $\epsilon$ ChaA} - This association is defined by Mamajek, Lawson \& Feigelson (2000).
We propose new members, enlarging it.
The well defined distance found by us indicates it is in front of the Cha complex.
As in YSSA, the northern border being in the Sacy's limits, we can not
be sure if this group is not a part of Lower Cen-Crux subgroup.

{\it e) $\beta$ PicA} - As described by Zuckerman et al.(2001) this association
is very close to the Sun, containing members with distances between 10
and 50 pc.
We proposed new members, someones at larger distances,
among them V4046 Sgr, a notorious object,
classified before as an isolated SB Classical T Tauri star.

{\it f) OctA} - Is a very homogeneous small group of almost aligned stars
(all young G stars)  near the South Celestial Pole.
As this region belongs to a completely surveyed area of the SACY,
new members have too be found by other means.

{\it g) AnA} - Another small group, very concentrated in velocity but
spread out in space.
Having only two stars with measured parallaxes,
it should be viewed with caution.

\acknowledgments
This work was partially supported by a CNPq - Brazil grant to C. A. O. Torres
(pr. 200356/02-0)

\end{document}